\documentclass[aps,prl,reprint]{revtex4-1}
\usepackage{blindtext}
\usepackage{braket}
\usepackage{bm}
\usepackage{color}
\usepackage{amsmath,amssymb}
\usepackage[legacycolonsymbols]{mathtools} 
\usepackage{mathrsfs} 
\usepackage{caption}
\usepackage{graphicx}
\usepackage{subcaption} 

\def\c{\hat{\psi}^\dagger(\bm{x})}
\def\a{\hat{\psi}(\bm{x})}
\def\x{(\bm{x},t)}
\def\d#1{\partial_{#1}}

\begin{document}
\title{Definition of current density in non-Hermitian quantum sytems}
\author{Hiroto Oka}

\affiliation{Division of Physics and Astronomy, Graduate School of Science, Kyoto University, Kyoto Unversity,
Kyoto 606-8502, Japan
Yukawa Institute for Theoretical Physics, Kyoto Unversity, Kyoto 606-8502, Japan}

\begin{abstract}
In recent years, non-Hermitian quantum systems (NHQS) have been actively studied. 
In conventional quantum mechanics, Hermiticity is a fundamental property of Hamiltonians. In NHQS, however, states evolve under non-Hermitian Hamiltonians and novel physical phenomena are predicted due to the non-Hermiticity. One difference from Hermitian systems is that the continuity equation (CE) does not hold in NHQS even when the number of particles is conserved. In this study, I extended the definition of current density so that CE holds also in NHQS. The newly defined current density does not only satisfy CE but also has physical meanings in the sense that it affects physical observables in the same manner as conventional current density.
\end{abstract}

\maketitle
\section{I. Introduction}
In recent years, non-Hermitian quantum systems (NHQS) where the time evolution is described by a non-Hermitian Hamiltonian have been attracting much attention \cite{ashida2020non,bender2007making,zhang2022review,okuma2023non}. Due to its non-Hermiticity, various novel phenomena are predicted \cite{gong2018topological,ding2022non,budich2020non}. There are several cases where NHQS appear. One of them is \textit{PT}-symmetric systems, where Hamiltonians are not Hermitian, while \textit{PT}-symmetric \cite{el2018non}. This field attracts much attention as extension of Hermitian quantum mechanics.

Another is the case where quantum mechanics itself has Hermiticity, but the time evolution of the system can be effectively described by a non-Hermitian Hamiltonian. In this research, I considered this type of NHQS, or NHQS in the quantum trajectory more specifically \cite{ashida2020non,minganti2020hybrid,minganti2019quantum,meden2023pt}. I present how NHQS appear in the quantum trajectory in the next section.

There are many deffereces in NHQS from Hermitian cases. One of them is that the continuity equation (CE) does not hold in NHQS generally even in U(1)-symmetric systems \cite{yan2024transport,takane2023probability,spourdalakis2016generalized}. 
In Hermitian particle systems, if the number of particles is conserved, in the other words, if the system has global U(1) gauge symmetry, CE holds between the particle density and the current density as follows.
\begin{align}
\partial_t \rho(\bm{x},t)=-\nabla \cdot \bm{j}(\bm{x},t)
\end{align}
In NHQS, however, EC does not hold generally if the current density $\bm{j}$ of the Hermitian case is used without modification. 
\begin{align}
\partial_t \rho(\bm{x},t) \neq -\nabla \cdot \bm{j}(\bm{x},t)
\end{align}
Therefore, I thought to extent the definition of current density so that CE also holds in NHQS. This is natural taking into account that current density is originally defined to satisfy CE in Hermitian systems. There is some previous research about the violation of CE in NHQS. In those research, non-Hermiticity is thought to correspond to gain and loss of particles, and this problem is resolved by introducing current density incoming to and outgoing from the system. However, as is discussed later, the non-Hermiticity in the quantum trajectory arises from quantum measurement (wave function collapse) and the time evolution by the non-Hermitian Hamiltonian is obtained by selecting only cases where no particles enter and leave the system. 
Based on this fact, I introduced not incoming and outgoing current density but current density which flows in the system so that EC holds in NHQS. 

In this study, I considerd non-Hermitain particles systems with U(1)-symmetry and  defined current density of the particles. For this purpose, I introduced an interaction between the particles and an electromagnetic field. In brief, I calculated the effect of the non-Hermiticity on the electromagnetic field and derived the current density of the particles indirectly. This means that this study focuses on current density of particles which have electric charge, or electric current density. Although this paper focuses on U(1) gauge field and electric current denisty, the method to define current density in NHQS can be extended to non-Abelian gauge fields and current density of corresponding charge (See appendix).

This paper is organized as follows. In this section, I introduce NHQS that appear in the quantum trajectory and non-Hermitian models considered in this study. Then, I present the violation of CE and explain the motivation of this research in detail. After that, in Section I\hspace{-1.2pt}I, I explain the method to define current density in NHQS and show the main result of this study.  Finally, in Section I\hspace{-1.2pt}I\hspace{-1.2pt}I, I present the conclusion and discussion.

In the following, operators corresponding to physical quantities are denoted with a hat, and when written without a hat, they represent the expectation values of the corresponding operators.
For examples, $\hat{\rho}(\bm{x})$ represents particle density operator at position $\bm{x}$ and 
$\rho(\bm{x},t)$ represents its expectation value for the state at time $t$.

Throughout this paper I use natural units with $\hbar=c=1$.
\subsection{Non-Hermitian quantum systems in the quantum trajectory}

Here, I give an overview of non-Hermitian quantum systems that appears in the quantum trajectory \cite{ashida2020non,minganti2020hybrid,minganti2019quantum,meden2023pt}.
This type of NHQS appears when the system interacts with measuring devices. Consider a composite system consisting of a system of interest and a measuring device. Hilbert spaces of the system and the measuring device are denoted by $\mathscr{H}_S$ and $\mathscr{H}_M$, respectively. The initial state is a product state $\ket{\psi}\bigotimes\ket{\chi}$, where $\ket{\psi}\in\mathscr{H}_S$ and $\ket{\chi}\in\mathscr{H}_M$.

The Hamiltonian of the system of interest and the interaction between the system and the measurement device are denoted by $\hat{H}_0$ and $\hat{H}_{int}$. 
Here, $\hat{H}_{int}$ is assumed to be the following form.

\begin{equation}
\hat{H}_{int}=\sum_{m}g(\hat{A}_m\otimes \hat{B}_m+\hat{A}^\dagger_m\otimes \hat{B}^\dagger_m )
\end{equation}
where $g$ is a parameter of the strength of the interaction. $\hat{B}_m$ is an operator that acts on $\mathscr{H}_M$ and transitions $\ket{\chi}$ to $\ket{\chi_m}$ correspondong to measurement outcome $m\,(m=1,2,3...)$. Furthermore $\hat{B}_m$ is required to satisfy
\begin{align}
&\hat{P}_m\hat{B}_{m'}=\delta_{m,m'}\hat{B}_{m'}\label{Hint condition}\\
&\Big ( \hat{P}_m=\ket{\chi_m}\bra{\chi_m}\Big)\notag\\
&\langle\chi_m|\chi\rangle=0\notag\\
&\langle\chi_m|\chi_{m'}\rangle=\delta_{m,m'}\notag
\end{align} 
Therefore, the interaction $\hat{H}_{int}$ satisfies
\begin{align}
\label{int con}\bra{\chi}\hat{H}_{int}\ket{\chi}=0
\end{align}
Now an approximation is adopted. 
Consider the time evolution of the initial state over small time $\tau$. For this small time $\tau$ and the parameter $g$, it is assumed that $\tau<<1$, so $\tau^2$ without $g$ can be neglected. However, $g^2\tau$ is finite, so $ g^2\tau^2 \sim \tau $ remains.
Under this approximation, the time evolution of the composite system over the small time $\tau$ is

\begin{align}
\bra{\chi}\hat{H}_{int}^2\ket{\chi}=\sum_m \hat{L}_m^\dagger\hat{L}_m\\
\hat{L}_m=\sqrt{g^2\tau\bra{\chi}\hat{B}_m^\dagger\hat{B}_m\ket{\chi}}\hat{A}_m
\end {align}
After the time evolution for the small time $\tau$, we perform an measurement and extract the case where the state of the measuring device remains $\ket{\chi}$. This corresponds to applying the projection operator $\ket{\chi}\bra{\chi}$. This operation, which extracts only a certain measurement result, is called post-selection.
The post-selected state of the composite system is, therefore,
\begin{align}
\ket{\psi}\ket{\chi} 
\to&\ket{\chi}\bra{\chi}\{1-i\hat{H}_{S}\tau +\frac{(-i)^2}{2}\hat{H}_{int}^2\tau^2 \}\ket{\psi}\otimes\ket{\chi} \notag \\
=&(1-i\hat{H}\tau)\ket{\psi}\otimes\ket{\chi}
\end{align}
where
\begin{align}
\hat{H}&=\hat{H}_0-\frac{i}{2}\bra{\chi}\hat{H}^2_{int}\tau\ket{\chi}\notag \\
&=\hat{H}_0-\frac{i}{2}\sum_m\hat{L}_m^\dagger \hat{L}_m\label{nH Hamiltonian}
\end{align}
Here Eq. (\ref{Hint condition}) was used. Repeating measurement and post-selection every $\tau$, the time evolution of the system of interest can be described by the effective non-Hermitian Hamiltonian $\hat{H}$.
\begin{align}
&i\partial_t\ket{\psi(t)}=\hat{H}\ket{\psi(t)}\label{st time evo}
\end{align}
This is an overview of NHQS in the quantum trajectory. As showed above, the non-Hermiticity arises from measurements and post-selection. The effect of quantum measurements, or wave function collapse, is incorporated as the non-Hermiticity of the system of interest. Although changes of the state by measurements every $\tau$ are discontinuous, the state evolves continuously under the non-Hermitian Hamiltonian within the above approximation that $\tau$ is sufficiently small.

\subsection{Conservation of particle number}
In this study, I considered non-Hermitian particle systems that conserves the number of particles in the system of interest \cite{hamazaki2019non,lee2020many,kawabata2022many,gliozzi2024many}.
That is, if the particle number operator of the system of interest is denoted by $\hat{N}$, then
\begin{align}
[\hat{H},\hat{N}]=[\hat{H}^\dagger,\hat{N}]=0 \label{N pre}
\end{align}
where
\begin{align}
\hat{N} = \int \c\a d\bm{x}
\end{align}
and $\c,\a$ are creation and annihilation operators of the particle at position $\bm{x}$.

This type of non-Hermitian system appears even if the interaction between the systems of interest and the measuring devices does not conserve $\hat{N}$, due to post-selection.
Consider a Hamiltonian of a system of interest and an interaction with a measuring device like below.

\begin{align}
\hat{H}_0=\int_{\bm{x}} \c\frac{(-i\nabla)^2}{2m}\a+V(\bm{x})\c\a d\bm{x}
\end{align}
\begin{align}
\hat{H}_{int}=g \sum_m\int_{\bm{x}}\a \otimes \hat{B}_m +\c \otimes \hat{B}_m^\dagger d\bm{x}
\end{align}
While $\hat{H}_0$ conserves $\hat{N}$, on the other hand, $\hat{H}_{int}$ does not.
However, the non-Hermitian Hamiltonian obtained from this interaction does conserve $\hat{N}$.
\begin{align}
[\hat{H}_0+i\hat{\Gamma} , \hat{N}]=0
\end{align}
where
\begin{align}
&\hat{\Gamma}=\int \hat{\psi}^\dagger(\bm{x}) \Gamma_{\bm{x},\bm{y}} \hat{\psi}(\bm{y})d\bm{x} d{\bm{y}}\notag \\
&\Gamma_{\bm{x},\bm{y}}=\Gamma_{\bm{y},\bm{x}}^*\notag
\end{align}

Therefore, if the initial state is an eigenstate of $\hat{N}$, then the state evolved under this non-Hermitian Hamiltonian is also an eigenstate of $\hat{N}$ with the same eigenvalue.
\begin{align}
\hat{N}e^{-i\hat{H}t}\ket{\psi}&=Ne^{-i\hat{H}t}\ket{\psi}\\
\text{if} \, \,\hat{N}\ket{\psi}&=N\ket{\psi}\notag
\end{align}

This is due to post-selection. By performing post-selection, only cases are exracted where the number of particles in the system of interest does not change (see Fig.1). 
Including the above example, I considered non-Hermitian particle systems that conserves the number of particles.
\begin{figure}[htbp]
\begin{center}
\includegraphics[scale=0.4]{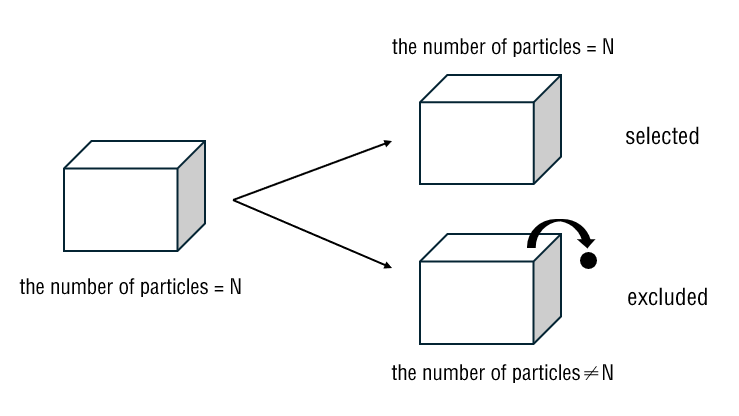}
\caption{\raggedright \small{Although the interaction can change the number of particles in the system of interest, only cases where it is conserved are extracted by post-selection.}}
\end{center}
\end{figure}

\subsection{Violation of the continuity equation and current density flowing in the system}
In this section, I show the violation of CE in NHQS and the reasons why I introduced current density flowing in the system of interest. Consider a non-Hermitian particle system that conserves the number of particles.
As you know, if $\hat{\Gamma}=0$, which means Hermitian cases, we can obtain current density $\bm{j}(\bm{x})$ which satisfies CE by calculating as follows.

\begin{align}\partial_t\rho(\bm{x},t)&=\partial_t \bra{\psi(t)}\c\a\ket{\psi(t)} \notag \\
&=\bra{\psi(t)}i[\hat{H}_0, \a\c]\ket{\psi(t)}\notag \\
&=-\nabla\cdot\bra{\psi(t)}\hat{j}(\bm{x})\ket{\psi(t)} \notag \\
&=-\nabla\cdot\bm{j}(\bm{x},t)\label{EOC}
\end{align} 
where $\hat{\rho}(\bm{x})=\c\a$ is particle density operator at position $\bm{x}$ and $\ket{\psi(t)}$ is the state at time $t$.

In NHQS, however, this $\bm{j}(\bm{x})$ does not satisfy CE due to the non-Hermitian part even when $\hat{H}$ conserves the number of particles \cite{yan2024transport,takane2023probability,spourdalakis2016generalized}.

\begin{align}\partial_t\rho(\bm{x},t)&= -\nabla\cdot\bm{j}(\bm{x},t) +\delta
\end{align}
where $\delta$ is a non-Hermiticity-induced term. 

In general, this problem is thought to arise from gain and loss of particles based on the fact that non-Hermitian Hamiltonians were originally introduced as models for systems with dissipation \cite{feshbach1958unified,gamow1928zur,breit1936capture}. 
In previous research, therefore, it is common to introduce current density incoming into and outgoing from the system to deal with this problem \cite{yan2024transport,takane2023probability}.
On the other hand, in this research, I introduced current density flowing in the system to resolve this problem (see Fig.2). 
\begin{figure}[htbp]
  \centering
  \begin{subfigure}{0.2\textwidth}
    \centering
    \includegraphics[scale=0.4]{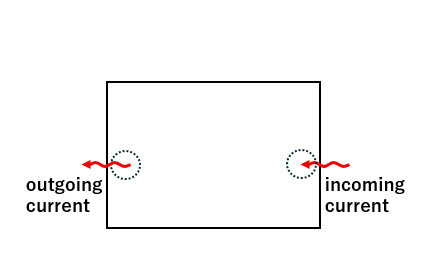}
    \caption{previous research}
  \end{subfigure}
  \hfill
  \begin{subfigure}{0.2\textwidth}
    \centering
    \includegraphics[scale=0.4]{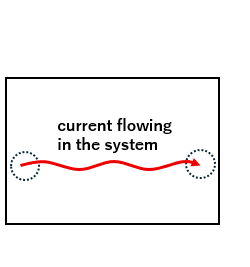}
    \caption{this research}
  \end{subfigure}
  \caption{\raggedright \small{In this research, instead of incoming and outgoing current density, current density flowing in the system was introduced so that CE holds in NHQS. }}
\end{figure}

I explain the reason in detail here. As mentioned above, non-Hermiticity is the result of the effect of measurements performed every short time $\tau$. This effect is discontinuous in time, but under the approximation that $\tau$ is sufficiently small, the time evolution can be described as continuous by a non-Hermitian Hamiltonian. Moreover, during such a small time $\tau$, the net number of particles entering and leaving the system of interest is zero thanks to post-selection. Under these conditons, only time evolutions without particle exchange is allowed. From these observations, it seems unnatural to consider gain and loss of particles. As a clearer example illustrating that non-Hermiticity does not correspond to particle gain and loss, consider an interaction $\hat{H}_{int}$ like below.    
\begin{align}
\hat{H}_{int}=g \int_{\bm{x}}\c\a \otimes (\hat{B}_m +\hat{B}_m^\dagger ) d\bm{x}
\end{align}
Also in NHQS derived from this type of interactions, the violation of CE can occur. However, this type of interactions apparently does not involve any particle exchange. Thus, it is not acceptable that the violation of CE arises from particle gain and loss. Taking these into account, instead of incoming and outgoing current density, I introduced current density flowing inside the system of interest so that CE holds.

In the following, only the system of interest (hereafter simply referred to as “the system”) is described and the space of the measuring device is omitted for simplicity. Note that interactions introduced later when defining current density is not interactions between the system and the measuring device but interactions within the system.

Here, I show the notation and the definition of expectation values used in this study.
The non-Hermitian Hamiltonian of the system is denoted by $\hat{H}$ using the original Hamiltonian of the system $\hat{H}_0$ and the non-Hermitian part $i\hat{\Gamma}$.
\begin{align}
\hat{H}=\hat{H}_0+ i\hat{\Gamma}
\end{align}
Be careful that both $\hat{H}_0$ and $\hat{\Gamma}$ are Hermitian.

And in this study, definitions of observables and their expectation values are the same as Hermitian case \cite{meden2023pt}. For the state $\ket{\psi(t)}$, the expectation value of an observable $\hat{O}$ is defined by
\begin{align}
O=\bra{\psi_N(t)}\hat{O} \ket{\psi_N(t)}=\frac{\bra{\psi(t)}\hat{O}\ket{\psi(t)}}{\langle\psi(t)|\psi(t)\rangle}
\end{align}
where
\begin{align}
\ket{\psi_N(t)}=\frac{\ket{\psi(t)}} {\sqrt{   \langle \psi(t) | \psi(t) \rangle  }    }
 \end{align}
Due to the non-Hermiticity, the norm ${\langle\psi(t)|\psi(t)\rangle}$ can change in time. However, from probabilistic interpretation of quantum mechanics, the norm of physical states should be always conserved. Thus, the state must be normalized at each time and expectation values need to be calculated with normalized states $\ket{\psi_N(t)}$.

For this reason, the expressions of the time evolution of expectation values are a little complicated in NHQS. Then, I applied a technique to $\hat{\Gamma}$ to simplify the expressions.
The technique is to redefine $\hat{\Gamma}$ at each time as below.
\begin{align}
\hat{\Gamma}\to \hat{\Gamma}-\bra{\psi(t)}\hat{\Gamma}\ket{\psi(t)} \notag
\end{align}
This technique ensures that the norm is conserved, therefore, expectation values can be expressed in the same way as Hermitian cases. 
\begin{align}
O(t)=\bra{\psi(t)}\hat{O}\ket{\psi(t)} =\braket{\hat{O}}
\end{align}
Hereafter, I denote expectation values $\bra{\psi(t)}\hat{O}\ket{\psi(t)} $ by $\braket{\hat{O}}$.
In addtion, owing to this technique, the time evolution of expectation values can be expressed without writing the norm change term as 
\begin{align}
\partial_t O(t) =\braket{ i[\hat{H}_0, \hat{O}] + \{ \hat{\Gamma} , \hat{O} \} }\label{ex time evo}
\end{align}
Note that this is merely a difference in expression and the result is the same.
Hereafter, although not explicitly stated, $\Gamma$ has the time-dependent c-number term and 
\begin{align}
\braket{\Gamma}=0 
\end{align}
holds at each time.

\section{I\hspace{-1.2pt}I. the method to define current density in NHQS}
Here, I explain the method to define current density in NHQS. As discussed previously, the purpose of this study is to define current density which flows in the system to satisfie CE in non-Hermitian particle systems with conserved particle number. However, it is impossible to define such current density only from the requirement of CE because countless current density can satisfy CE. Moreover, if one of them is defined as current density, its physical meaning is unclear. Then, I introduced an interaction between the particles and an electromagnetic filed and calculated Maxwell's equations in NHQS to define current density. I explain this in detail below.

As you know, in a Hermitian case which includes particles with charge $q$ and an electromagnetic field, Maxwell's equations are derived as \cite{weinberg2005quantum,kugo1979local}
\begin{align}
\nabla\cdot\bm{E}\x&=q \rho\x \label{Gauss} \\
\nabla\times\bm{B}\x&=q \bm{j}\x+\partial_t\bm{E}\x \label{amp}\\
\nabla \times \bm{E}\x&=-\partial_t \bm{B}\x \label{faraday}\\
\nabla \cdot \bm{B}\x&=0\label{no monopole}
\end{align}
where the quantities $\rho$ and $\bm{j}$ denote the particle density and current density, respectively. $\bm{E}$ and $\bm{B}$ represent the electric and magnetic field and are written with the scalar potential $\phi$ and the vector potential $\bm{A}$ as below.
\begin{align}
&\bm{E}=-\d{t} \bm{A}-\nabla\phi \label{electric field}\\
&\bm{B}=\nabla \times \bm{A}
\end{align}
As mentioned before, they are expectation values of corresponding operators for the state $\ket{\psi(t)}$.

What is important here is that CE can be derived from equations (\ref{Gauss}) and (\ref{amp}). To put it differently, when CE does not hold in NHQS, conventional Maxwell's equations above do not hold and must be modified in some way. 
Based on this fact, I calculated modified Maxwell's equations in NHQS and defined current density from them. 
Note that this method needs electric charge and interactions. Then, this study focuses on electric current density of particles. 
Although this study focuses on electromagnetic field and electric current density, this method can be extended to non-Abelian gauge theory and corresponding gauge current density (See Appendix). 

Defining current density indirectly from Maxwell's equations may look strange. But, physical properties, such as diamagnetism in the Meissner effect, are determined not solely by current density but in combination with Maxwell's equations \cite{london1935electromagnetic,bardeen1957theory}. In this sense, defining current density in this way is justified because this current density determines physical properties.

\subsection{Maxwell's equations and current density in NHQS}

In this section, I present the detailed method for defining current density in NHQS, which is the main result of this research. Consider a Hamiltonian consisting of particles with charge $q$ and an electromagnetic field in the Coulomb gauge ($\nabla \cdot \hat{\bm{A}}=0$).
\begin{align}
&\hat{H}_0=\hat{H}_{part}+\hat{H}_{Rad}+\hat{H}_{Coul}\\
&\hat{H}_{part}=\int \c \frac{(-i\nabla-q\hat{\bm{A}}(\bm{x}))^2  }{2m} \a\, d\bm{x}\\
&\hat{H}_{Rad}=\int \frac{1}{2}\big(\hat{\bm{E}}_T^2(\bm{x})+\hat{\bm{B}}^2(\bm{x})\big )\, d\bm{x} \\
&\hat{H}_{Coul}=\int q\hat{\rho}(\bm{x})\hat{\phi}(\bm{x}) \, d\bm{x} 
\end{align}
where $\hat{H}_{part}$, $\hat{H}_{Rad}$ and $\hat{H}_{Coul}$ correspond to charged particles, a radiation field and the Coulomb interaction respectively. $\hat{\bm{E}}_T$ and $\hat{\bm{B}}$ denote the operators of the transvers electric filed and magnetic field. And $\hat{\bm{A}}$, $\hat{\phi}$ denote the electromagnetic potentials.
The particle may be either a boson or a fermion; in either case, the conventional current density derived from the particle Hamiltonian $\hat{H}_{part}$ is denoted by $\bm{j}$. And the non-Hermitian part is denoted by $i\hat{\Gamma}$. This $\bm{j}$ does not satisfy CE when the non-Hermitian part $i\hat{\Gamma}\neq 0$. 
 
Note that definitions of operators are the same as the Hermitian case and the relations between the operators below still hold.
\begin{align}
\hat{\phi}(\bm{x})&=\int \frac{q}{4\pi}\frac{\hat{\rho}(\bm{x}')}{|\bm{x}-\bm{x}'|}d\bm{x}' \label{phi and rho}\\
\hat{\bm{B}}&=\nabla \times \hat{\bm{A}} \label{B and A}\\
\hat{\bm{E}}&=-i[\hat{H}_0, \hat{\bm{A}}]-\nabla \hat{\phi}\label{Eop}
\end{align}
where $\bm{E}$ is the electric field operator.
Here, I adopted the Schr\"odinger picture and the Coulomb gauge.
Gauge field theory is generally formulated in the Heisenberg picture. However, for NHQS, the Heisenberg picture becomes complicated \cite{ju2025heisenberg} and I adopted the Schr\"odinger picture in this study. 
As for gauge choise, since NHQS are derived from gauge invariant systems, so are they and the result does not depend on gauge choice. See Appendix for other gauges. 

Now I proceed to present how Maxwell's equations  are modified in NHQS.
Initially, I mention the relationship between the electric field and the electromagnetic potentials in NHQS.
Eq.(\ref{Eop}) leads the conventional relation between them in Hermitian case.
\begin{align}
\bm{E}=-\partial_t \bm{A} -\nabla \phi 
\end{align}
In NHQS, however, it is slightly modified as below because the time evolution of expectation values are given by Eq.(\ref{ex time evo}).  
\begin{align}
\bm{E}=-\partial_t \bm{A} +\braket { \{  \hat{\Gamma} , \hat{\bm{A}}  \} } -\nabla \phi \label{new electric field}
\end{align}
Next, I show how Eq.(\ref{Gauss}) (\ref{no monopole}) and (\ref{amp}) are modified in NHQS.
First, Eq.(\ref{Gauss}) and (\ref{no monopole}) still hold in NHQS because Eq.(\ref{phi and rho}) and (\ref{B and A}) are the same as the Hermitian case. 
\begin{align}
\nabla\cdot \bm{E}&=q \rho \\
\nabla \cdot \bm{B}&=0
\end{align}
On the other hand, Eq.(\ref{amp}) is modified in NHQS from Eq.(\ref{ex time evo}) as below.
\begin{align}
\nabla \times \bm{B}= q \bm{j}+\partial_t \bm{E}-\braket{ \{ \hat{\Gamma} , \hat{\bm{E}} \} }
\end{align}
Here, I define a quantity $\bm{E}'$ as
\begin{align}
\bm{E}'(\bm{x},t)=\bm{E}(\bm{x},t)-\braket{ \{  \hat{\Gamma},\hat{\bm{A}_T} \} } \label{E' and E}
\end{align}
where $\hat{\bm{A}}_T$ is the transverse part of the vector potential. 
With Eq.(\ref{new electric field}), $\bm{E}'$ can be written as
\begin{align}
\bm{E}'=-\partial_t \bm{A}-\nabla\phi + \braket{ \{ \hat{\Gamma}, \hat{\bm{A}}_L\} }
\end{align}
where $\hat{\bm{A}}_L$ is the longitudinal part of the vector potential.  
Now, Coulom gauge is chosen, then  
\begin{align}
\bm{E}'=-\partial_t \bm{A}-\nabla\phi
\end{align}
Putting these equations together, Maxwell's equations in NHQS are written  with $\bm{E}'$ as below.
\begin{align}
\nabla\cdot\bm{E}'\x&=q \rho\x \label{new Gauss} \\
\nabla\times\bm{B}\x&=q \widetilde{\bm{j}}\x+\partial_t\bm{E}'\x \label{new amp}\\
\nabla \times \bm{E}'\x&=-\partial_t \bm{B}\x \label{faraday}\\
\nabla \cdot \bm{B}\x&=0\label{new no monopole}
\end{align}
where
\begin{align}
\widetilde{\bm{j}}&=\bm{j}-\frac{1}{q} \big \{  \braket{ \{ \hat{\Gamma} , \hat{\bm{E}} \} }-\partial_t \braket{\{ \hat{\Gamma}, \hat{\bm{A}}_T \} } \big\} \label{new current}\\
&=\bm{j}+\delta \bm{j}\notag
\end{align}
$\bm{j}$ is conventioanl current density from the Hermitian Hamiltonian $\hat{H}_0$ and $\delta\bm{j}$ is the non-Hermiticity-induced current density.

This $\widetilde{\bm{j}}$ is precisely the current density newly defined in this study.
From Eq. (\ref{new Gauss}) and (\ref{new amp}), $\widetilde{\bm{j}}$ satisfies CE, that is to say, the time variation of particle density is given by the divergence of $\widetilde{\bm{j}}$.
\begin{align}
\partial_t \rho = -\nabla \cdot \widetilde{\bm{j}}
\end{align}
Moreover,  $\widetilde{\bm{j}}$ does not only satisfy CE mathematically but also has physical meanings. From Maxwell's equations in NHQS, the vector potential can be written as
\begin{align}
\bm{A}\x=&\frac{1}{4\pi}\int\frac{q\widetilde{\bm{j}}_T(t',\bm{x}')}{|\bm{x}-\bm{x}'|}d\bm{x}'\\
t'=&t-|\bm{x}-\bm{x}'| \notag
\end{align}
$\widetilde{\bm{j}}_{T}$ is the transverse part of $\widetilde{\bm{j}}$.
This expression is exactly the same form as the equation where current density constructs vector potential in Coulomb gauge, except that $\bm{j}$ is replaced by $\widetilde{\bm{j}}$.
As you can see, the vector potential is governed not by $\bm{j}$ but by $\widetilde{\bm{j}}$. In other words, in NHQS, it is $\widetilde{\bm{j}}$ that plays the role of current density with respect to electromagnetic potentials.
Of course, $\bm{E}'$ is different from the electric field $\bm{E}$ because of the term $\braket { \{ \hat{\Gamma} , \hat{\bm{A}}_T \} }$ as in Eq.(\ref{E' and E}). However, in regions where the electromagnetic field behaves classically, this term is zero and 
\begin{align}
\bm{E}=\bm{E}'&=-\partial_t \bm{A} -\nabla \phi \\
\because \braket { \{ \hat{\Gamma} , \hat{\bm{A}}_T \} }& =2\bm{A}_T\braket{ \hat{\Gamma} }=0\notag
\end{align}
Therefore, if we measure the electromagnetic field in such regions, it is precisely $\widetilde{\bm{j}}$ that behaves as current density with respect to the electromagnetic field, or the measured quantites.   

The reason why I consider measurements in classical electromagnetic filed region is that the state is stable under measurements of it. If the state is affected by measurements, the time evolution of the state is no longer described by the non-Hermitian Hamiltonian and out of scope of this study. Thus, in this study, which consider current denisity in NHQS, measurements of classical electromagnetic field are assumed. 

Although the electromagnetic field should be classical in the measured region, in the region where current density flows, the electromagnetic field needs to be quantum and entangled with the non-Hermitian term in order for the non-Hermiticity-induced current density to arise. If not, it cannot arise because the following two terms in Eq.(\ref{new current}) are both vanish.
\begin{align}
\braket{\{ \hat{\Gamma}, \bm{A}\} }=\braket{\{ \hat{\Gamma}, \bm{E} \} }=0
\end{align}
In summary, if we measure classical electromagnetic field, the measured quantities are governed by newly defined current density with the effect of the non-Hermiticity-induced current density flowing in quantum electromagnetic field region (see Fig.3).
\begin{figure}[htbp]
\begin{center}
\includegraphics[scale=0.5]{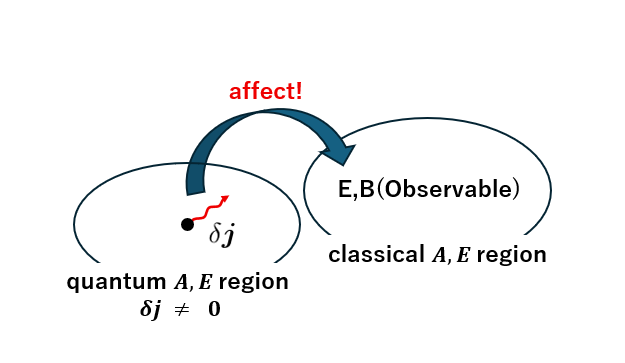}
\caption{\raggedright \small{Measured quantities, or classical EM field are affected by the non-Hermiticity-induced current density flowing in quantum EM filed region.} }
\end{center}
\end{figure}
\section{I\hspace{-1.2pt}I\hspace{-1.2pt}I. conclusion and discussion}
In this study, I defined current density in non-Hermitian particle systems based on the fact that CE does not hold with the conventional current density in NHQS. The difference from previous research is that current density flowing in the systems is considered in this study. In previous research, the violation of CE is thought to arise from gain and loss of particles and current density incoming to and outgoing from the system is considered. However, in this study, motivated by the fact that net particle exchange is prohibited under post-selections, and moreover, that the violation of CE can occur even in systems which inherently lack particle exchange, I considered current density flowing in the system. To define such current density, I introduced interactions between the particles and an electromagnetic filed and calculated Maxwell's equations in NHQS. Instead of deriving current density directly from the Hamiltonian, I calculated the effect of the non-Hermiticity on the electromagnetic field and defined current density indirectly from it.
As a result, newly defined current density satisfies CE and moreover, it behaves the same way as the conventional current density toward electromagnetic potentials and field. Note that this method does not show apparently that such non-Hermiticity-induced current density really exists. However, what is important here is that measured quantities behave as if such current density exists. 

As repeatedly mentioned, non-Hermiticity and the violation of CE originate from measurements and wave function collapse, rather than from particle gain and loss.
Of course, this effect is discontinuous in time. However, within an approximation where continuous description by a non-Hermitian Hamiltonian is valid, current density corresponding to the effect can be obtained (See Fig.4).
\begin{figure}[htbp]
\begin{center}
\includegraphics[scale=0.5]{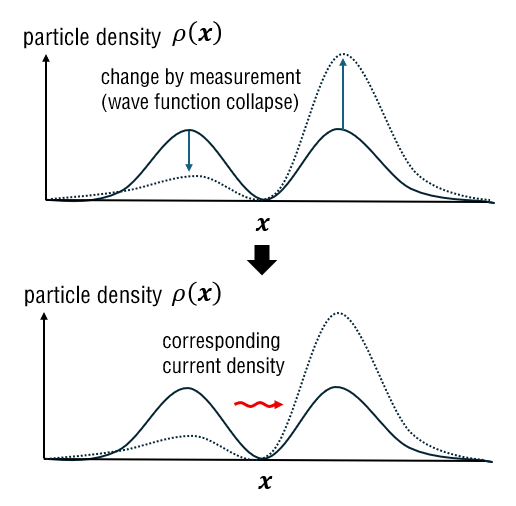}
\caption{\raggedright \small{Current density corresponding to discontinuous particle density change by wave function collapse. solid and dotted line: particle density before and after a measurement.} }
\end{center}
\end{figure}

The continuous description by non-Hermitian Hamiltonians is valid under the approximation which ignores second and higher orders in $\tau$. On the other hand, the non-Hermiticity-induced current density is of the order of $g^2 \tau$ and this corresponds to the zeroth order in $\tau$. Therefore, even when higher-order terms of $\tau$ are taken into account, the effect of the current density is still considered to remain. In future research, I expect that novel physical properties originating from this current density will be observed.
\section{Acknowledgement}
This work was supported by JST SPRING, Grant
Number JPMJSP2110.

\bibliographystyle{unsrt} 
\bibliography{myref}

\section{Appendix} 
In the main text, the Coulomb gauge is adopted to define current density in NHQS. Here, I present a derivation of the same current density with other gauges. Becchi-Rouet-Stora-Tyutin (BRST) formalism is used here and the original Hamiltonian $\hat{H}_0$ is \cite{kugo1979local}
\begin{align}
\hat{H}_0=\hat{H}_{part}&+\int \! d\bm{x} \,\frac{1}{2}( \hat{\bm{E}}^2 + \hat{\bm{B}}^2 )
+ \phi \,\nabla \cdot \hat{\bm{E}} \notag \\
&+ \hat{b} (-\,\nabla \cdot \hat{\bm{A}} + F) 
- \frac{\alpha}{2}\,\hat{b}^{2}+ \hat{\pi}_{c}\,\hat{\pi}_{\bar c}
+ \nabla \hat{\bar c} \cdot \nabla \hat{c}
\end{align}
$\hat{H}_{part}$ is the particle Hamiltonian which includes the interaction between the particles and the electromagnetic field. 
\begin{align}
\hat{H}_{part}
=\int \hat{\psi}^{\dagger}
\left[
\frac{1}{2m}\left(-i\nabla - q\,\hat{\bm{A}}\right)^{2}
+ q\,\phi
\right]\hat{\psi} \,d\bm{x}
\end{align}
$\hat{b}$ denotes the Nakanishi–Lautrup auxiliary field, while $\hat{c}$ and $\hat{\bar{c}}$ are the Faddeev–Popov ghost and anti-ghost fields, respectively. $\hat{\pi}_c$ and $\hat{\pi}_{\bar c}$ are canonical momentum of the ghost and anti-ghost field. And $F$ is a continuous function which decides the gauge condition.
This Hamiltonian has BRST-symmetry and commutates with BRST charge operator $\hat{Q}_{BRST}$.
\begin{align}
[\hat{H}_0, \hat{Q}_{BRST}]=0
\end{align}
In BRST formalism, physical states are defined by BRST charge as below. 
\begin{align}
\hat{Q}_{\small{BRST}}\ket{\psi_{phy}}=0\label{physical condition}
\end{align}
In the Hermitian case, with this condition used, $\hat{H}_0$ leads to Maxwell's equations (\ref{Gauss})-(\ref{no monopole}) under the following gauge condtion.
\begin{align}
\partial_t \phi +\nabla \cdot \bm{A}= F
\end{align}
Next, trun to the non-Hermitian case. Even in NHQS, I still assumed the physical state condition (\ref{physical condition}), that is to say, even if the time evolution is governed by a non-Hermitian Hamiltonian, the state should be always physical and satisfie Eq.(\ref{physical condition}).
In fact, if the interaction $\hat{H}_{int}$ introduced in Section I is BRST-symmetric, the non-Hermtian Hamiltonian derived from it is aslo BRST-symetric from Eq.(\ref{nH Hamiltonian}). Thus, if the original theory is BRST-symmetric and the state is initially physical, the state which evolevs under the non-Hermitian Hamiltonian remains physical over time. I performed the calculations under this assumption.

Within the Hermitian BRST formalism, Gauss's law follows from the fact that it can be expressed as a anti-commutator of the BRST charge, together with the condition that the states are physical \cite{kugo1979local,lee1993brst}. 
\begin{align}
\nabla \cdot \bm{E}- q\rho=\braket{\nabla\cdot\hat{\bm{E}}-q\hat{\rho}}=\braket{i\{\hat{Q}_{BRST}, \hat{\pi}_c\}}=0
\end{align}
Because definitions of observables and the physical condtion  are the same as the Hermitian case, Gauss's law still holds in NHQS.
As the case with the Coulomb gauge, by introducing $\bm{E}'$ and calculating the time evolution of the expectation values, Maxwell's equations in NHQS with this gauge are derived as follows.
\begin{align}
\nabla \cdot \bm{E}'&=q \rho\\
\nabla\times\bm{B}&=q \widetilde{\bm{j}}+\partial_t\bm{E}'\\
\nabla \times \bm{E}'&=-\d{t} \bm{B} \\
\nabla\cdot \bm{B}&=0
\end{align}
where
\begin{align}
\bm{E}'&=-\partial_t \bm{A}-\nabla\phi + \braket{ \{ \hat{\Gamma}, \hat{\bm{A}}_L\} }\\
\bm{B}&=\nabla \times \bm{A}\\
\widetilde{\bm{j}}&=\bm{j}-\frac{1}{q} \big \{  \braket{ \{ \hat{\Gamma} , \hat{\bm{E}} \} }-\partial_t \braket{\{ \hat{\Gamma}, \hat{\bm{A}}_T \} } \big\}
\end{align}
Note that the gauge condition is slightly different from the Hermitian case ($\hat{\Gamma}=0$) from Eq.(\ref{ex time evo}). 
\begin{align}
\partial_t \phi +\nabla \cdot \bm{A}= F+\braket{\{ \hat{\Gamma} , \phi\} }
\end{align}
Define $\phi'$ as
\begin{align}
\phi'&= \phi+ \delta\phi\\
\nabla \delta \phi &= - \braket{\{ \hat{\Gamma} , \hat{\bm{A}}_L \} } 
\end{align}
With this $\phi'$, Maxwell's equations in NHQS are expressed as below.
\begin{align}
&\nabla \cdot (-\partial_t \bm{A}-\nabla \phi')= q\rho \\
\nabla\times &\nabla \times  \bm{A}=q\widetilde{\bm{j}}+\partial_t (-\partial_t\bm{A} -\nabla \phi') \\
&\qquad \bm{E}'=-\partial_t \bm{A}-\phi' \\
&\qquad \bm{B}=\nabla \times \bm{A}
\end{align}
with
\begin{align}
\partial_t\phi' +\nabla\cdot \bm{A}= F+\braket{\{ \hat{\Gamma} , \phi \}}+\partial_t \delta\phi
\end{align}
As mentioned in the main text, in classical electromagnetic field region, $\bm{E}'=\bm{E}$ and $\widetilde{\bm{j}}$ behaves as current density with respect to the electromagnetic field.
In this way, it turns out that the quantity $\widetilde{\bm{j}}$ can be defined as current density in this gauge, as well as in the Coulomb gauge. Defining current density using BRST formalism has a benefit. Because BRST formalism includes not only U(1) gauge but also non-Abelian gauge, therefore, this method can be extended to non-Abelian gauge fields and corresponding charge current density.

\end{document}